\documentclass[12pt]{article}
\usepackage{axodraw}
\usepackage{epsfig}
\usepackage{latexsym}
\usepackage{amssymb}
\usepackage{amsfonts}
\newcommand{\be}{\begin{equation}}
\newcommand{\ee}{\end{equation}}
\newcommand{\bea}{\begin{eqnarray}}
\newcommand{\eea}{\end{eqnarray}}

\newcommand{\gm}{\gamma}
\newcommand{\Gm}{\Gamma}

\newcommand{\ep}{\varepsilon}
\newcommand{\sg}{\sigma}

\newcommand{\pa}{\partial}
\newcommand{\dd}{\mbox{d}}

\newcommand{\nn}{\nonumber}
\newcommand{\D}{\mbox{d}}
\newcommand{\lra}{\leftrightarrow}

\newcommand{\SA}{\left\{ \begin{array}{ll}}
\newcommand{\Sa}{\left[ \begin{array}{ll}}
\newcommand{\FA}{\end{array}\right.}

\newcommand{\leftdot}{$\!\!\!\mbox{\bf{.}}\,\,$}
\newtheorem{tr}{Theorem}
\newcommand{\BTh}{\begin{tr}\leftdot}
\newcommand{\ETh}{\end{tr}}
\newtheorem{lmm}{Lemma}
\newcommand{\BLm}{\begin{lmm}\leftdot}
\newcommand{\ELm}{\end{lmm}}
\newtheorem{deff}{Definition}
\newcommand{\BDf}{\begin{deff}\leftdot}
\newcommand{\EDf}{\end{deff}}
\newtheorem{stttt}{Step}
\newcommand{\BStt}{\begin{stttt}\leftdot}
\newcommand{\EStt}{\end{stttt}}
\newenvironment{proof}{$\blacktriangleleft$ }{$\blacktriangleright$
\indent}
\newcommand{\BPr}{\begin{proof}}
\newcommand{\EPr}{\end{proof}}
\newcommand{\BZM}{\begin{zamet}}
\newcommand{\EZM}{\end{zamet}}

\newcommand{\p}[1]{\langle #1\rangle}

\newcommand{\mm}{\mathfrak{m}}

\newcommand{\pp}{\mathfrak{p}}

\newcommand{\I}{i}

\def\p{{\bf 1}^+}
\def\pp{{\bf 2}^+}

\def\m{{\bf 1}^-}
\def\mm{{\bf 2}^-}

\begin{document}
\parindent=1.5pc

\begin{titlepage}
%\rightline{September 2005}

\bigskip
\begin{center}
{{\large\bf
Applying Gr\"obner Bases to Solve Reduction Problems\\ for Feynman Integrals
} \\
\vglue 5pt
\vglue 1.0cm
{\large  A.V. Smirnov}\footnote{E-mail: asmirnov@rdm.ru}\\
\baselineskip=14pt
\vspace{2mm}
{\normalsize Mechanical and Mathematical Department and\\
Scientific Research Computer Center of
Moscow State University
   }\\
\baselineskip=14pt
\vspace{2mm}
and\\
\baselineskip=14pt
\vspace{2mm}
{\large   V.A. Smirnov}\footnote{E-mail: vsmirnov@mail.desy.de}\\
\baselineskip=14pt
\vspace{2mm}
{\normalsize
Nuclear Physics Institute of Moscow State University\\
%Moscow 119992, Russia
}
\baselineskip=14pt
\vspace{2mm}
\vglue 0.8cm
{Abstract}}
\end{center}
\vglue 0.3cm
{\rightskip=3pc
 \leftskip=3pc
\noindent
We describe how Gr\"obner bases can be used to solve the reduction
problem for Feynman integrals, i.e. to construct an algorithm that
provides the possibility to express a Feynman integral of a given
family as a linear combination of some master integrals.
Our approach is based on a generalized Buchberger algorithm for
constructing Gr\"obner-type bases associated with polynomials of
shift operators. We illustrate it through various examples of
reduction problems for families of one- and two-loop Feynman
integrals. We also solve the reduction problem for a family of
integrals contributing to the three-loop static quark potential.
\vglue 0.8cm}
\end{titlepage}

\section{Introduction}

The important mathematical problem of evaluating Feynman integrals
arises naturally in elementary-particle physics when one
treats quantum-theoretical amplitudes in the framework of perturbation
theory.
This problem originated in the early days of perturbative quantum field
theory. Over more than five decades, a great variety of methods
for evaluating Feynman integrals has been developed.
However, to check whether the Standard Model
or its extensions describe adequately
particle interactions observed in experiments, one needs to perform more and
more sophisticated calculations, so that one tries not only to
update existing methods but also develop new effective methods of
evaluating Feynman integrals.

After a tensor reduction based on some projectors a
given Feynman graph generates various scalar Feynman integrals
that have the same structure of the integrand with various
distributions of powers of propagators which we shall also call
{\em indices}.
%Now let us consider the situation that arises when calculating
%Feynman integrals.
Let $F(a_1,a_2,\ldots,a_n)$ be a scalar
dimensionally regularized \cite{dimreg} {}Feynman integral
corresponding to a
given graph and labelled by the (integer) indices, $a_i$:
%\be
%F(a_1,a_2,\ldots)=
%\int\ldots\int \frac{\dd^d k_1 \dd^d k_2\ldots}{(p_1^2-m_1^2)^{a_1}
%(p_2^2-m_2^2)^{a_2}\ldots}\;.
%\label{FI}
%\ee
\bea
  F(a_1,\ldots,a_n) &=&
  \int \cdots \int \frac{\dd^d k_1\ldots \dd^d k_h}
  {E_1^{a_1}\ldots E_n^{a_n}}\,,
  \label{eqbn}
\eea
where $k_i$, $i=1,\ldots,h$, are loop momenta
and the denominators are given by
\bea
  E_{r}&=&\sum_{i\geq j \geq 1} A^{i j}_{r} \, p_i \cdot p_j
- m_{r}^2 \,,
  \label{denom}
\eea
with  $r=1,\ldots,n$.
The matrix $A^{i j}_{r}$ depends on the choice of the loop momenta.
The momenta $p_i$ are either
the loop momenta $p_i=k_i, \; i=1,\ldots,h$, or
independent external momenta
$p_{h+1}=q_1,\ldots,p_{h+n}=q_n$ of the graph.
Irreducible polynomials in the numerator can be represented as
denominators raised to negative powers.
For example, the denominator corresponding to the propagator of a
massless particle is $k^2=k_0^2-\vec{k}^2$.
Usual prescriptions $k^2=k^2+i 0$, etc. are implied.
Formally, dimensional regularization \cite{dimreg}
is denoted by the change
$\dd^4 k=\dd k_0 \dd \vec{k}\to \dd^d k$, where
$d=4-2\ep$ is a general complex number.
The Feynman integrals are functions of the masses, $m_i$,
and kinematic invariants, $q_i\cdot q_j$. However, we shall
omit this dependence because we shall pay special attention to
the dependence on the indices. We shall also omit the dependence
on $d$.

A straightforward
strategy is to evaluate, by some methods, every scalar Feynman
integral resulting from the given graph.
If the number of these integrals is small this strategy is quite
reasonable. In non-trivial situations, where the number of
different scalar integrals can be at the level of hundreds and thousands,
this strategy looks too complicated.
%can be hardly realized in practice.
A well-known
optimal strategy here is to derive, without calculation, and then
apply some relations between the given family of Feynman integrals
as {\it recurrence relations}.
A well-known standard way to obtain such relations is provided
by the method of integration by parts
(IBP) \cite{IBP}
which is based on the fact that any dimensionally regularized integral
of the form
\be
\int \D^d k_1 \D^d k_2\ldots \frac{\pa f}{\pa k_i^{\mu}}
\label{IBP-RR}
\ee
is equal to zero.
Here $f$ is the integrand in
(\ref{eqbn}).
More precisely, one tries to use IBP
relations in order to express a general dimensionally
regularized integral of the
given family as a linear combination of some
`irreducible' integrals which
are also called {\it master} integrals.
Therefore the whole problem decomposes into two parts:
the construction of a reduction algorithm and the evaluation of
the master Feynman integrals.

There were several recent attempts to make the reduction
procedure systematic:

({\em i}) Using the fact that the total number of IBP equations
grows faster than the number of independent Feynman integrals, when one
increases the total power of the numerator and denominator, one can
sooner or later obtain an overdetermined system
of equations \cite{LGR1,LGR2} which can be solved.
(There is a public version of implementing the corresponding
algorithm on a computer \cite{AnLa}.)

({\em ii}) Using relations that can be obtained by
tricks with shifting dimension \cite{Tar}.

({\em iii}) Baikov's method \cite{Bai}.

Another attempt in this direction is based on the use of Gr\"obner bases
\cite{Buch}.
The first attempt to apply the theory of Gr\"obner bases in
the reduction problems for Feynman integrals was made in
\cite{Tar1}, where IBP relations were reduced to differential
equations.
To do this, it is assumed that there is a non-zero mass for each
line. %The typical combination $a_i {\bf i}^+$, where  ${\bf i}^+=Y_i$
%is the shift operator, is then naturally transformed into
%the operator of differentiation in the corresponding mass.
For differential equations one can then apply some standard
algorithms for constructing corresponding Gr\"obner bases.
%Still in the end it is necessary to get rid of the masses
%if one wants to deal with typical Feynman integrals that arise in
%practical calculations.

In \cite{Tar1,Gerdt} it was pointed out that the straightforward
implementation of the Buchberger algorithm in the case of IBP
relations is problematic because it requires
a lot of computer time even in simple examples.
One of the possible modifications is related to the Janet bases
\cite{Gerdt05}.
%However, this approach works only with the data given by the IBP
%relations and neglects the physical specifics of the problem.
We are going to modify the Buchberger algorithm in
another way, taking into account explicitly such properties  as
boundary conditions (which characterize all the regions
of indices where the Feynman integrals are equal to zero),
so that it will be possible to apply it to
%solve IBP relations
solve the reduction problem in complicated situations.

In the next section, we shall briefly describe what the
Gr\"obner basis and the Buchberger algorithm are in the classical
problem related to solving systems of algebraic equations. In Section~3,
we shall turn to IBP relations and describe our strategy of
constructing Gr\"obner bases associated with the given problem,
with the help of a modification of the standard
Buchberger algorithm. We shall explain how these results can be
applied to solve IBP relations.
We shall illustrate our strategy, through various examples, in
Section~4. In particular, we shall apply our algorithm to
a family of three-loop Feynman integrals with a one-loop
insertion relevant to the three-loop quark potential. We shall
also evaluate the master integrals using the method
based on Mellin--Barnes representation. In Conclusion, we shall
characterize the status and perspectives of our method.

\section{Gr\"obner basis and Buchberger algorithm}

The notion of the Gr\"obner basis was invented by Buchberger
\cite{Buch} when
he constructed an algorithm to answer certain questions on the
structure of ideals of polynomial rings.

Let ${\cal A}=\mathbb C[x_1,\ldots,x_n]$ be
the commutative ring of polynomials of $n$ variables $x_1,\ldots,x_n$
over $\mathbb C$ and $\cal I\subset \cal A$ be an
ideal\footnote{
%A ring is a set with two operations: multiplication and addition.
A non-empty subset $\cal I$ of a ring $R$
is called a left (right) ideal if ({\em i}) for any $a,b\in \cal I$ one has $a+b\in
\cal I$ and ({\em ii}) for any $a\in {\cal I}, c\in R$ one has
%$a c\in I$ and
$c a\in \cal I$ ($a c\in \cal I$ respectively).
In the case of commutative rings there is no difference
between left and right ideals.}. A classical problem is to
construct an algorithm that shows whether a given element $g\in
{\cal A}$ is a member of $\cal I$ or not. A finite set of polynomials
in $\cal I$ is said to be a \textit{basis} of $\cal I$ if any
element of $\cal I$ can be represented as a linear combination of its
elements, where the coefficients are some elements of $\cal A$.
Let us fix a basis $\{f_1,f_2,\ldots,f_k\}$ of $\cal I$. The
problem is   %, being given a polynomial $g\in {\cal A}$,
to find out whether
there are polynomials $r_1,\ldots,r_k \in {\cal A}$ such that $g=
r_1 f_1+\ldots+r_k f_k$.

%In this classical case, a solution to this problem can be given by
%an algorithm consisting of a {\em finite} sequence of steps. This
%algorithm is called Buchberger algorithm. It is based in
%constructing a Gr\"obner basis.

Let $n=1$. In this case any ideal is generated by one
element $f=a_0+a_1 x+a_2 x^2 +\ldots + a_m x^m$.
Now if we want to find out whether an element
$g=b_0+b_1 x+b_2 x^2 +\ldots + b_l x^l$ can
be represented as $rf$ we first check if $l\geq m$. If so, we replace
$g$ with $g-(b_l/a_m)x^{l-m}f$, `killing' the leading term of
$g$. This procedure is repeated until
the degree of a `current' polynomial obtained from $g$ becomes
less than $m$.
It is clear that the resulting polynomial is equal to zero if and only if $g$ can
be represented as $rf$.

Now let $n>1$. Let us consider an algorithm that will answer this problem for
some bases of the ideal. (We will see later that
this problem can be solved if we have a so-called Gr\"obner basis at hand.)
To describe it, one needs the notion of
an {\em ordering of monomials} $c x_1^{i_1}\ldots x_n^{i_n}$
where $c\in\mathbb C$ and the
notion of the {\em leading term} (an analogue of the intuitive one in the case $n=1$).
%An ordering on the set of monomials $\cal M$ is
%a map $\cal M=\{x_1^{i_1}\ldots x_n^{i_n}\}\righttarrow \mathbb C$ such that
%it depends only on the \textit{degree}
%of a monom
In the simplest variant of {\em
lexicographical} ordering, a set $(i_1,\ldots,i_n)$ is said to be
{\em higher} than a set $(j_1,\ldots,j_n)$ if there is $l\leq n$ such
that $i_1=j_1$, $i_2=j_2$, \ldots, $i_{l-1}=j_{l-1}$ and
$i_l>j_l$. The ordering is denoted as
$(i_1,\ldots,i_n)\succ (j_1,\ldots,j_n)$.
%\succ \prec
We shall also say that the corresponding monomial
$c x_1^{i_1}\ldots x_n^{i_n}$ is higher than the monomial
$c' x_1^{j_1}\ldots x_n^{j_n}$.

One can introduce various orderings, for example,
%one can compare, first, the sums of powers $i_1,\ldots,i_n$.
the {\em\ degree-lexicographical} ordering,
where $(i_1,\ldots,i_n)\succ (j_1,\ldots,j_n)$
if $\sum i_k >   \sum j_k$, or $\sum i_k =   \sum j_k$
and $(i_1,\ldots,i_n)\succ (j_1,\ldots,j_n)$ in the sense of
the lexicographical ordering.
The only two axioms that the ordering
%on ${\cal M}=\{x_1^{i_1}\ldots x_n^{i_n}\}$
has to satisfy are
that $1$ is the only minimal element under this ordering and
that if $f_1\succ f_2$ then $gf_1\succ gf_2$ for any $g$.

Let us fix an ordering. The {\em leading term}
(under this ordering) of a polynomial
\[
P(x_1,\ldots,x_n)=\sum c_{i_1,\ldots,i_n} x_1^{i_1}\ldots  x_n^{i_n}
\]
is the non-zero monomial
$c_{i_1^0,\ldots,i_n^0} x_1^{i_1^0}\ldots  x_n^{i_n^0}$ such that
the degree $(i_1^0,\ldots,i_n^0)$ is higher than the degrees
of other monomials in $P$.
Let us denote it by $\hat{P}$. We have $P=\hat{P}+\tilde{P}$,
 where $\tilde{P}$ is the
sum of the remaining terms.

Let us return to the problem formulated above. Suppose that the
leading  term of the given polynomial $g$ is divisible by the
leading  term or some polynomial of the basis, i.e.
$\hat{g}=Q \hat{f_i}$ where $Q$ is a monomial. Let $g_1=g- Q f_i$.
It is clear that the leading  term of $g_1$ is lower than the
leading  term of $g$ and that $g_1\in {\cal I}$ if and
only if $g\in {\cal I}$.
One can go further and proceed with $g_1$ as with $g$,
using the same $f_i$ or some other element $f_j$ of the initial
basis, and obtain similarly $g_2,g_3,\ldots$. The
procedure is repeated until one obtains $g_l\equiv 0$
or an element $g_l$ such that $\hat{g_l}$ is not divisible by any
leading  term $\hat{f_i}$. We will say that $g$ is
reduced to $g_l$ modulo the basis $\{f_1,f_2,\ldots,f_k\}$.

%After it is impossible to make such a reduction with the help of
%the element $f_1$ one can continue the procedure using $f_2$ as
%well as other elements $f_3,\ldots,f_k$.

A basis $\{f_1,f_2,\ldots,f_k\}$ is called a {\em Gr\"obner basis} of the
given ideal if any polynomial $g\in \cal I$ is reduced by the described
procedure to zero for any sequence of reductions.
Given a Gr\"obner basis we obtain an algorithm
to verify whether an element $g\in {\cal A}$ is a member of $\cal I$.
There are many other questions on the structure of the ideal that
can be answered constructively if one has a Gr\"obner basis, but
they are beyond the topic of the paper.

Generally a basis is not a Gr\"obner basis.
Let $f_1=x_1$ and $f_2=1+x_2^2$ and let $\cal I$ be generated by
$f_1$ and $f_2$. It is easy to verify that $\{f_1,f_2\}$ is a
Gr\"obner basis of $\cal I$. Now let $f'_1=x_1 x_2$. The set
$\{f'_1,f_2\}$ is again a basis of $\cal I$ (indeed,
$f'_1=x_2 f_1$ and $f_1= -x_2 f'_1 + x_1 f_2$). However,
$\{f'_1,f_2\}$
is not a Gr\"obner basis because the element $x_1\in\cal I$ cannot be
reduced modulo $\{f'_1,f_2\}$.

On the other hand, given any initial
basis  $\{f_1,f_2,\ldots,f_k\}$ of the ideal $\cal I$
% is not a Gr\"obner basis
one can construct a Gr\"obner basis starting from it and
using the {\em Buchberger algorithm} which consists of the following
steps.

Suppose that $\hat{f_i}=w q_i$ and  $\hat{f_j}=w q_j$ where
$w, q_i$ and $q_j$ are monomials and $w$ is not a constant.
Define $f_{i,j}= f_i q_j-
f_j q_i$. Reduce this polynomial modulo the set $\{f_i\}$ as described
above. If one obtains a non-zero polynomial by this reduction,
add it to the set  $\{f_i\}$. Consider then the other elements with
$\hat{f'_i}=w q'_i$ and  $\hat{f'_j}=w' q'_j$
for some non-constant $w'$.
If there is nothing to do according to this
procedure one obtains a Gr\"obner basis. It has been proven by
Buchberger \cite{Buch} that such a procedure stops after a finite number of steps.

The Buchberger algorithm can take much computer time to construct
a Gr\"obner basis, but once it has been constructed, one can use the reduction
procedure which works generally much faster.
%Moreover, the same
%bases can be used to answer several other questions on the
%structure of the ideal.

To conclude, the problem formulated in the beginning of this
section can be solved by choosing an ordering and constructing the
corresponding Gr\"obner basis using the Buchberger algorithm.
After that, one applies the reduction procedure modulo the constructed
Gr\"obner basis to verify whether a given element belongs to the
given ideal $\cal I$.

%\vspace{1cm}
% We are going to describe a certain algorithm
%that starts with $g_1=g$ and constructs a finite sequence of
%elements $g_1,g_2,\ldots,g_k$. In the case where $g_k=0$ if and only
%if $g_1\in I$, we will call the basis $\{f_1,f_2\ldots,f_i\}$ a
%Gr\"obner basis. Therefore, provided one has constructed a Gr\"obner
%bases of the ideal, checking whether an element is a member of
%this ideal turns into an algorithmically solved problem. It is
%also worth noting that a few other important questions about the
%structure of the ideal can easily be solved provided one has a Gr\"obner
%basis.
%
%The next important point is that, in the classical case, there
%exists an algorithm constructing a Gr\"obner basis for any ideal,
%called the Buchberger algorithm.

\section{Reduction problem for Feynman integrals}

Practically, one uses relations (\ref{IBP-RR}) of the following
form:
\be
\int\ldots\int \dd^d k_1 \dd^d k_2\ldots
\frac{\pa}{\pa k_i}\left( p_j
\frac{1}{E_1^{a_1}\ldots E_N^{a_N}}
\right)   =0\;.
\label{FI-IBP}
\ee
Here $E_r$ are denominators in (\ref{eqbn}),
$k_1 ,\ldots,k_h$ are loop momenta and $p_1=k_1,\ldots,
p_h=k_h, p_{h+1}=q_1,\dots, p_{h+n}=q_n$, where $q_1,\ldots,q_n$
are independent external momenta.

After the differentiation, resulting scalar products,
$k_i\cdot k_j$ and $k_i\cdot q_j$
are expressed in terms of the factors in the denominator,
by inverting (\ref{denom}), and one arrives at
IBP relations which can be written as %take the form
%
%There is a function $F(a_1,a_2,\ldots,a_n)$ and
%a set of relations of the sort
\begin{equation}
\sum c_i F(a_1+b_{i,1},\ldots,a_n+b_{i,n})
%\equiv f_i \cdot F(a_1 ,\ldots,a_n)
=0\,,
\label{IBP}
\end{equation}
%\noindent where $F is
where $b_{i,j}$ are integer,
$c_i$ are polynomials in $a_j$,
$d$, masses $m_i$ and kinematic invariants,
%and the arguments of
and $F(a_1,\ldots,a_n)$ are Feynman integrals (\ref{eqbn}) of the
given family.
%are shifted by integer numbers
These relations can be written in terms of shift operators
${\bf i}^+$ and ${\bf i}^-$ which are defined as
\[{\bf i}^{\pm} \cdot
F(a_1,a_2,\ldots,a_n)=F(a_1,\ldots,a_{i-1},a_i\pm
1,a_{i+1},\ldots,a_n)\,.\]

At this point, we would like to turn from the `physical'
shift operators ${\bf i}^{\pm}$  to
`mathematical' shift operators. (We believe that the physical notation
can be ambiguous: for example, it is not immediately clear whether
the operators are applied to a function of the indices, or
to some of its values.)

Let $\mathcal K$ be the field of rational functions of physical
variables $m_i$, $q_i\cdot q_j$, $d$, and $\cal A$ be the
algebra\footnote{An algebra over a field is a vector space over
this field and a ring at the same time.} over $\mathcal K$
generated by elements $Y_i$, $Y^{-1}_i$ and $A_i$ with the
following relations: \bea Y_i Y_j &=& Y_j Y_i,\;\;\;\ A_i A_j =
A_j A_i, \;\;\;Y_i A_j = A_j Y_i+\delta_{i,j}Y_i,
\\ \nn
Y^-_i Y^-_j &=& Y^-_j Y^-_i,\;\;\;
Y^-_i Y_j = Y_j Y^-_i, \;\;\;
Y^-_i A_j = A_j Y^-_i-\delta_{i,j}Y_i,\;\;\;Y^-_i Y_i=1\,
\eea
where $\delta_{i,j}=1$ if $i=j$ and $0$ otherwise.
For convenience we will write $(Y^-_i)^k=Y^{-k}_i$.
Let $\mathcal F$ be the field of functions of $n$ integer
arguments $a_1,a_2,\ldots,a_n$.
The algebra $\cal A$
acts on this field\footnote {
({\em i}) for any $a\in\cal A$ and $f\in \cal F$
we have an element $a\cdot f\in\cal F$;
({\em ii}) for any $a,b\in\cal A$ and $f,g\in \cal F$ we have
$(a+b)\cdot(f+g)=a\cdot f+a\cdot g+b\cdot f+b\cdot g$;
({\em iii}) for any $a,b\in\cal A$ and $f\in \cal F$ we have
$(ab)\cdot f=a\cdot(b\cdot f)$.}, where
\bea
(Y_i\cdot F)(a_1,a_2,\ldots,a_n)=F(a_1,\ldots,a_{i-1},a_i+
1,a_{i+1},\ldots,a_n)\,,
\\\nn
%operators $A^-_i=A^{-1}_i$
%$Y_i={\bf i}^+$
%$Y^-_i={\bf i}^+$
%and the operators of multiplication
(A_i \cdot F)(a_1,a_2,\ldots,a_n)=a_i F(a_1,a_2,\ldots,a_n)\,.
\eea
%The ring $\cal A$ can be considered as `almost'
%polynomial ring, where the coefficients generally do not commute
%with monomials.
% --- as ${\cal A}=(\mathbb C[a_1,\ldots,a_n])[x_1,\ldots,x_n]$.
%We have a noncommutative algebra with relations
%\[
%Y^{\pm}_i A_i =(A_i\pm 1) Y^{\pm}_i\,.
%\]

Let us turn back to the problem of calculating Feynman integrals.
The left-hand sides of relations (\ref{IBP}) can be represented
as elements of the ring
${\cal A}$ applied to $F$; we will denote these elements by $f_1,\ldots,f_n$.
Now, for $F(a_1,\ldots,a_n)$ defined by (\ref{eqbn}), we have
\be
f_i\cdot F=0\mbox{ or }(f_i\cdot F)(a_1,\ldots,a_n)=0
\ee
for all $i$. Let us
generate a (left) ideal $\cal I$ by the elements $f_1,\ldots,f_n$.
We will call $\cal I$ the \textit{ideal of the IBP relations}.
Obviously,
\be
f\cdot F=0\,, \;\; \mbox{or} \;\;
(f\cdot F)(a_1,\ldots,a_n)=0 \;\; \mbox{ for any }\;\;f\in {\cal I}\,.
\ee

Our goal is to express the value of $F$ at an arbitrary point
$(a_1,a_2,\ldots,a_n)$ in terms of the values of $F$ in a few
specially chosen points, i.e. master integrals.
This problem can be solved similarly
to the algebraic problem described in Section~2.
Consider, for example, the case, where all the indices $a_i$ are
positive. Then one has
\be
F(a_1,a_2,\ldots,a_n)
=(Y_1^{a_1-1}\ldots Y_n^{a_n-1}\cdot F)(1,1,\ldots,1)\,.
\ee
%can obtain????? $F(a_1,a_2,\ldots,a_n)$ as the
%monomial $Y_1^{a_1-1}\ldots Y_n^{a_n-1}$ acting on
%$F(1,1,\ldots,1)$, or another monomial acting on a Feynman
%integral as some other chosen point.
The idea of the method is to
reduce the monomial $Y_1^{a_1-1}\ldots Y_n^{a_n-1}$ modulo the ideal
of IBP relations.
Let us consider a trivial example of such a situation.

\textit{Example~1}. One-loop vacuum massive Feynman integrals
\be
 F(a) = \int \frac{\dd^d k }{(k^2-m^2)^{a}}
  \; .
\label{ex51}
\ee
%In this chapter, we are concentrating on the dependence of Feynman integrals
%on the powers of the propagators so that we will usually omit dependence
%on dimension, masses and external momenta.
Let us forget that these integrals can be evaluated explicitly, in
terms of gamma functions.
%try to exploit information following from IBP.
The IBP identity
\be
  \int \dd^d k \frac{\pa}{\pa k} \cdot k \frac{1}{(k^2-m^2)^{a}}=0
  \; ,
\label{ex51f1}
\ee
%with $(\pa/(\pa k))\cdo k=(\pa /(\pa k_{\mu})) k_{\mu}$,
leads to the relation
\be
(d-2a+2) F(a-1)-2 (a-1) m^2 F(a)=0
  \; .
\label{ex51f11}
\ee
%which immediately can be written in a recursive way
%since the integral with a given $a$ can be expressed in terms
%of the integral with $a-1$ for $a\neq 1$.
We see that any Feynman integral $F(a)$ where $a>1$ can be
expressed recursively in terms of one integral $F(1)\equiv I_1$
which we therefore qualify as a master integral.
(Observe that all the integrals with non-positive integer indices
are integrals without scale and
are naturally put to zero within dimensional regularization.)
%This can be done explicitly here:
%\be
%F(a)=\frac{(-1)^a \left(1-d/2\right)_{a-1}}{(a-1)! (m^2)^{a-1}}I_1
%  \; ,
%\label{ex51f3}
%\ee
%where $\left(x\right)_{a}$ is the Pochhammer symbol and the only master integral is
%\be
% I_1 =-\I\pi^{d/2} \Gm(1-d/2) (m^2)^{d/2-1}
%  \; .
%\label{ex51f4}
%\ee
%***************
%Let us present a trivial non-physical example of such a situation.
%Take $n=1$ and one equation $F(a+1)=(a+1)F(a)$.
%Our goal is to evaluate $F(a)$ for positive $a$ knowing $F(1)$ .
%Of course, this equation is already recursive with a simple
%solution, $F(a)=a!F(1)$ but let us

Let us demonstrate how the reduction procedure can lead to the same
result.
(We realize that this way is more complicated
in this simple situation. However, we will see later that
its generalization provides simplifications and enables us to
solve complicated problems.)
The IBP relation (\ref{ex51f11}) gives us one
element $f=2m^2 A Y-(d-2A)\in{\cal A}$
(the element $(f\cdot F)(a-1)$ is the left-hand side of (\ref{ex51f11})).
Set ${\cal I}={\cal A} f$
(for any $g\in \cal A$ and $F\in \cal F$ we have $g f\cdot F$=0).
%Now,
%we can evaluate $F(a)$ using the following manipulations:
We have
%if we want to calculate $F(a)$ we write
\bea
2m^2 (A + a - 2) Y^{a-1}=(2m^2 (A + a - 2) Y^{a-1}-Y^{a-2}f)+Y^{a-2}f
&&  \nn \\ &&  \hspace*{-110mm}
=(2m^2 (A + a - 2) Y^{a-1}- Y^{a-2}(2m^2 A Y-(d-2A)))+Y^{a-2}f
\nn \\ &&  \hspace*{-110mm}
= Y^{a-2}(d-2A)+Y^{a-2}f=
(d-2A-2a+4)Y^{a-2}+Y^{a-2}f\,.
\eea
The relation
$2m^2 (A + a - 2) Y^{a-1}=(d-2A-2a+4)Y^{a-2}+X_1$,
where $X_1\in {\cal I}$,
represents one step of the reduction procedure.
If we stop the reduction at this point, we get
\bea
2m^2(a-1)F(a)=(2m^2 (a - 1)\cdot F)(a)
=(2m^2 (A + a - 2) Y^{a-1}\cdot F)(1)
&&  \nn \\ &&  \hspace*{-130mm}
= ((d-2A-2a+4)Y^{a-2})\cdot F)(1)+(Y^{a-2}f\cdot F)(1)
\nn \\ &&  \hspace*{-130mm}
= (d-2a+2) F(a-1)+((A-1)Y^{a-2}\cdot F)(1)= (d-2a+2) F(a-1)\,,
\eea
%
%
%2m^2(2a-2)F(a)=(2m^2 (A + a - 2)\cdot F)(a)
%=(2m^2 (A + a - 2) Y^{a-1}\cdot F)(1)
%&&  \nn \\ &&  \hspace*{-140mm}
%= ((d-2A-2a+4)Y^{a-2})\cdot F)(1)
%+(Y^{a-2}f\cdot F)(1)
%\nn \\ &&  \hspace*{-140mm}
%= (d-2a+2) F(a-1)+((A-1)Y^{a-2}\cdot F)(1)= (d-2a+2) F(a-1)\,,
i.e. the equation (\ref{ex51f11}) we started from. But moving further with the reduction modulo $\cal I$ we obtain
\be
P_1(A,a,m)Y^{a-1}=P_2(A,a,m)+X'\,,
\ee
where $P_1$ and $P_2$ are polynomials obtained during the
reduction and $X'\in {\cal I}$
(note that this algorithm is constructive and is realized as a computer code).
Now we can apply this equation to $F$ and take the value at $1$:
\bea
P_1(1,a,m)F(a)=(P_1(A,a,m)Y^{a-1}\cdot F)(1)
&&  \nn \\ &&  \hspace*{-55mm}
=((P_2(A,a,m)+X')\cdot F)(1)=P_2(1,a,m)F(1)\,.
\eea

It is enough to notice that $P_1$ is a product of
the leading  coefficients (the formal definition
in the non-commutative case will be given later)
of $f$, $Y f$ and so on.
Thus $P_1(1,a,m)$ is the product of the leading  coefficient
$2m^2 A$ of $f$ with $A$ replaced with all integers from $1$ to
$a-1$, hence non-zero. After dividing by this value we obtain the
needed representation.
%After applying the equality to $F$ and
%taking the value in $1$ we obtain a product of these highest
%coefficients taken
%
%represent $Y^{a-1}$ as a linear combination
%of a polynomial function of $A$ and an element of $I$.
%Applying this equality to $F(1)$ we obtain the known result.

Thus it looks tempting to generalize the standard reduction procedure
and reduce the monomial $Y_1^{a_1-1}\ldots Y_n^{a_n-1}$ so that
the resulting polynomial has a smaller degree in a certain sense. In
this case we would represent $F(a_1,a_2,\ldots,a_n)$ as a linear
combination of $F(a'_1,a'_2,\ldots,a'_n)$ for `smaller' $a'_i$.

This method works indeed, but first we need to introduce some
notation.
%We have $A^-_i A_i=A_i A^-_i=1$, therefore we can write
%$(A^-)^i=A^{-i}$.
We will say that an element $X\in \cal A$ is written in
the {\em proper} form if it is represented as
\be
X=\sum c_j(A_1,\ldots,A_n)\prod_i Y^{d_{i,j}}_i,
\ee
where $c_j$ are polynomials and $d_{i,j}$ are integers.
(So, all the operators $A_i$ are placed on the left from
the operators $Y_i$.)
Obviously any element $X\in\cal A$ has a unique proper
form. We will say that an element of $\cal A$
is a \textit{monomial} if in its proper form only one coefficient function $c_j$
is non-zero. We will say that the degree of a monomial
$c(A_1,\ldots,A_n)\prod_i Y^{d_{i}}_i$
is
$\{d_1,\ldots,d_n\}$.
We will say that a monomial
$c(A_1,\ldots,A_n)\prod_i Y^{d_{i}}_i$ is \textit{divisible}
by a monomial
$c'(A_1,\ldots,A_n)\prod_i Y^{d'_{i}}_i$
if $d'_i\geq d_i$ for all $i$.

Let us define a subalgebra $\cal A^+\subset \cal A$
generated as an algebra by $Y_i$ and $A_i$ (but not $Y^{-1}_i$)
and set ${\cal I}^+={\cal I}\cap {\cal A}^+$.
Obviously, ${\cal I}^+$ is an ideal in $\cal
A^+$ and ${\cal A}{\cal I}^+=\cal I$.
In the same way as in the classical
situation,
we introduce the notion of an ordering, leading  term, highest
degree of an element of $\cal A^+$ and the reduction modulo an ideal.
%and the
% defined almost in the same way.
The only problem is
that the leading coefficient of an element of $\cal A^+$
is now a polynomial function, so it does not generally have
an inverse element. Thus the reduction procedures lead us to
a relation
\be
c_0(A_1,\ldots,A_n) Y_1^{a_1-1}\ldots Y_n^{a_n-1}
=\sum_j c_j(A_1,\ldots,A_n)\prod_i Y^{d_{i,j}}_i+X',
\label{reduction}
\ee
where $X'\in {\cal I}$, $c_i$ are polynomials in $A_j$,
and none of the monomials on the right-hand side of the relation is
divisible by a leading  monomial of
an element of the basis ${\cal I}^+$.
Applying this equality to $F$ and taking the value at $(1,\ldots,1)$
we derive
\be
q_0 F(a_1,\ldots,a_n)=\sum_j q_j F(d_{1,j}+1,\ldots,d_{n,j}+1)\,,
\label{reductionapplied}
\ee
where $q_i$ do not depend on $a_j$. In the case where $q_0$ is non-zero
we can divide by it and obtain the desired representation.

As in the classical situation, one says that a finite set
$\{f_1,\ldots,f_k\}$ is a \textit{Gr\"obner basis} of an ideal
$\cal I$ if any element $f\in\cal I$ is reduced by this procedure
to zero. The number of different degrees $\{d_1,\ldots,d_n\}$
arising on the right-hand side of (\ref{reduction}) is
minimal possible if we have a Gr\"obner basis.
%Moreover, if it is known that all the values of $F(a_1,\ldots,a_n)$
%can be reduced to a finite number of values
%(and in all known physical examples this statement holds),
%then the reduction modulo the
%Gr\"obner basis also gives a finite
%number of different degrees.
%and we obtain a reduction of a given
%If the set of such residual monomials is infinite
%(this is the usual case), we do not have a solution of the
%reduction problem in the sense of reducing any given
%Feynman integral to a {\em finite} set of master integrals.
%In such a situation, one can try to
A Gr\"obner basis for
the ideal of IBP relations can be constructed
using the Buchberger algorithm
(a generalization of the algorithm explained
in Section~2).
%Then one can hope that the number of the master
%integrals will be finite.

%A straightforward generalization of the Buchberger algorithm works
%in a few simple examples, but makes the computer hopelessly freeze
%once you move to something nontrivial.
%Anyway, this reduction can be implemented as
%a finite algorithm, and one may hope that such a reduction will
%works faster than standard reduction procedures.
%If one has constructed an analogue of the
%Gr\"obner basis of the ideal $I$, this reduction can be implemented as an
%algorithm that often works faster than standard reduction
%procedures.
%However, the construction of Gr\"obner bases
%for polynomials of the shift operators turns out
%to be a complicated problem. In a few words, there exists an
%algorithm for this, but it turns to be completely inefficient and
%makes the computer freeze even in examples with small numbers of the propagators.
Let us now illustrate these points using a very simple

{\em Example~2.}
The family of one-loop massless propagator integrals
\be
F(a_1,a_2)  =
\int \frac{\dd^d k}{(k^2)^{a_1}((q-k)^{2})^{a_2}} \;.
\label{ex52}
\ee
%We shall always imply integer indices $a_i$ until Example~6.
%We omit the dependence on the external momentum squared, $q^2$ and
%the dimensional regularization parameter, $d$.
%In addition,
We have the boundary conditions,
$F(a_1,a_2)=0$ if $a_1 \leq 0$ or $a_2 \leq0$,
which correspond to putting to zero any integral without scale
within dimensional regularization.
As it is well known, this integral can be evaluated explicitly:
\be
F(a_1,a_2)  =
\I \pi^{d/2} \frac{(-1)^{a_1+a_2}\Gm(a_1+a_2+\ep-2)
\Gm(2-\ep-a_1)\Gm(2-\ep-a_2)}{(-q^2)^{a_1+a_2-2+\ep}
\Gm(a_1)\Gm(a_2)\Gm(4-a_1-a_2-2\ep)}
  \;,
\label{ex52-res}
\ee
but let us forget about this and consider the problem of the
reduction to master integrals.

The two IBP identities
\be
\int \dd^d k
\frac{\pa}{\pa k}\left(l
\frac{1}{(k^2)^{a_1}((q-k)^{2})^{a_2}}=0\right) \;,
\label{ex52-1}
\ee
with $l=k$ and $l=q$ give the following two IBP relations
\bea
d-2 a_1-a_2 -a_2 \pp (\m -q^2 )&=&0\;,
\label{fi-1-rr1}
\\
a_2-a_1- a_1\p (q^2 -\mm) -a_2 \pp (\m  -q^2)&=&0\;.
\label{fi-1-rr2}
\eea
%in the sense that they are applied to any integral
%$F(a_1,a_2)$.

These relations are defined by the elements
\bea
f_1&=&d-2 a_1-a_2 -a_2 Y_2   (Y_1^{-1} -q^2 ) \;,
\label{fi-1-rr1-op}
\\
f_2&=&(a_2-a_1)Y_1 Y_2- a_1 Y_1 (q^2 -Y_2^{-1}) -a_2 Y_2 (Y_1^{-1}  -q^2)
\;.
\label{fi-1-rr2-op}
\eea
which generate the ideal of IBP relations.
%Since we have boundary conditions it is natural to try to
%reduce a given integral to integrals with lowest indices
%and, therefore to turn to the operators $Y_1$ and $Y_2$ as basic
%operators. We do this by
Let us multiply (\ref{fi-1-rr1-op}) by $Y_1$
and (\ref{fi-1-rr2-op}) by $Y_1 Y_2$ to obtain
a basis of ${\cal I}^+\in {\cal A}^+$:
\bea
f'_1&=& (d  - 2 a_1 - a_2- 2)Y_1 -a_2 Y_2 (1-q^2 Y_1)
\;,
\label{fi-1-rr1-op2}
\\
f'_2&=& a_2-a_1-(a_1+1)Y_1^2 (q^2 Y_2-1 )
+( a_2+1 )Y_2^2 (q^2 Y_1-1 )
\;.
\label{fi-1-rr2-op2}
\eea

If we introduce  an ordering for polynomials in
the operators $Y_i$ and define the corresponding reduction procedure
modulo the operators $f'_1$ and $f'_2$ we shall obtain the
possibility to represent any given monomial as
\be
Y_1^{a_1-1} Y_2^{a_2-1}= r_1 f'_1 + r_2 f'_2
+ \sum c_{ij} Y_1^{i-1} Y_2^{j-1}\,,
\ee
where $r_1$ and $r_2$ are some elements of the ring $\cal A$.
So, if we act by this relation on
$F$ and take the value at $(1,1)$ we shall obtain
%and
%into account the validity of the IBP relations
%we shall obtain %(see also \cite{GeRo})
\be
F(a_1,a_2) =  \sum_{i,j} c_{ij} F(i,j) \,.
\label{red1}
\ee

We discover, however, that the set of integrals that
appear on the right-hand side of these relations obtained for
various $a_1$ and $a_2$ is infinite.
%The experience tells us, however that the number of master
%integrals is finite. To arrive at an algorithm which provides the
%finite number of the master integrals let us turn to the Gr\"obner
%basis.
%A specialization of
The Buchberger
algorithm described in Section~2 leads, with the degree-lexicographic
order, to the Gr\"obner basis
consisting of the following two elements:
\bea
g_1=2 a_1  Y_1  - d a_1  Y_1  + 2 a_1 ^2 Y_1  - 2 a_2 Y_2 + d a_2 Y_2 -
 2 a_2^2 Y_2
\label{gb11}
&&  \\ &&  \hspace*{-102mm}
g_2= 4 a_1  Y_1  - 4 d a_1  Y_1  + d^2 a_1  Y_1  + 8 a_1 ^2 Y_1  -
 4 d a_1 ^2 Y_1  + 4 a_1 ^3 Y_1  + 2 a_1  a_2 Y_1
\nn \\ &&  \hspace*{-102mm}
 - d a_1  a_2 Y_1
 + 2 a_1 ^2 a_2 Y_1  + 2 a_1  a_2 Y_2 - d a_1  a_2 Y_2 +
 2 a_1 ^2 a_2 Y_2 - 4 q^2 a_2 Y_2^2
\nn \\ &&  \hspace*{-102mm}
 + d q^2 a_2 Y_2^2 -
 6 q^2 a_2^2 Y_2^2 + d q^2 a_2^2 Y_2^2 - 2 q^2 a_2^3 Y_2^2\;.
\label{gb12}
\eea

Now, the reduction modulo these two elements provides only a finite
number of integrals in the corresponding relations (\ref{red1}).
In fact,
the degree of $g_1$ is $(1,0)$ and
the degree of $g_2$ is $(0,2)$, so we meet just the two integrals in this set,
$F(1, 1)$ and $F(1, 2)$, and call them
master integrals. For example, we have
\be
F(2, 3)=\frac{(d-8) (d-5)}{2q^2}F(1, 2)\,.
\ee
However, we do not obtain a connection of $F(1, 1)$ and $F(1, 2)$,
although we know, due to explicit solutions of the reduction
procedure, that they are connected. This is a disturbing point. Of
course, it is preferable to have only one master
integral in this trivial example, so that we are going to
develop an algorithm which reveals a minimal number of
the master integrals at least in simple examples.

Starting from example $3$, we will have at least one more
complication: the variables $a_i$, generally, can be not only
positive but also negative. (In the previous example, $a_1$ and
$a_2$ were positive due to the boundary conditions.)
Generally, we have to consider each variable $a_i$ to be either positive
or non-positive. Of course, for every family of Feynman integrals,
there will be some boundary conditions. (In particular, if all the
arguments $a_i$ are non-positive any Feynman integral is zero.)

Thus, if we have a family of Feynman integrals,
$F(a_1,\ldots,a_n)$,
we are going to consider each variable $a_i$ to satisfy $a_i>0$
or $a_i\leq 0$. Consequently, we have to consider $2^n$ regions
that we shall call {\em sectors} and label them by subsets
$\nu \subseteq \{1,\ldots,n\}$. The corresponding sector
$\sg_{\nu}$ is defined as $\{ (a_1,\ldots,a_n): a_i>0\;\;
\mbox{if}\;\; i\in \nu\,,\;\;
a_i\leq 0\;\;
\mbox{if}\;\; i \not\in \nu\}.$

In the sector where all $a_i$ are positive,
we considered the ring $\cal A^+\subset\cal A$
and the operators $Y_i$ as basic operators. (See the previous
example.) Quite similarly, in a given sector $\sg_{\nu}$ it is
natural to consider the subalgebra $\cal A^\nu\subset\cal A$ generated by
the operators $A_i$ and the operators $Y_i$ for $i\in \nu$ and
$Y_i^{-1}$ for other $i$. Within this definition we have
${\cal A}^{\{1,\ldots,n\}}=\cal A^+$.

Thus the first idea is to construct a Gr\"obner basis for each of
the $2^n$ sectors, or at least for all non-trivial sectors. (We
call a sector \textit{trivial} if all the given Feynman integrals are
identically zero in it due to boundary conditions.) This approach however
faces many problems:

\begin{enumerate}
\item Each of the non-trivial sectors will give us at least one
point where we have to evaluate $F$;
\item The number of points where we have to evaluate
$F$ in a given sector is generally greater than the real number of
master integrals (In the last example there is only one
master integral but %we get to evaluate the values
we obtain $F(1, 1)$ and $F(1, 2)$ after the reduction);
\item Even if one has constructed all the needed bases,
the reduction may fail in cases where the coefficient $q_0$
in eq.~(\ref{reductionapplied}) is zero. This problem
arises because all leading  coefficients are polynomials in $A_i$,
so that they can be equal to zero at certain points;
\item Although the method leads us theoretically to constructing a
Gr\"obner basis, all known practical implementations fail to work
even already in four-dimensional examples.
\end{enumerate}

%\textit{non-degenerate} sectors -
% for all of them

Therefore this specialization of the Buchberger
algorithm turns out to be completely impractical in sufficiently
complicated examples.
Our algorithm is a certain modification of the Buchberger algorithm.
%differs from the
%in two points.
%
%First, we are not obliged to construct
%Gr\"obner bases for all the sectors, but can restrict ourselves to
%what we call {\em sector bases}, or, briefly $s$-bases.
To characterize it we need to introduce some notation.

Let $\cal A^{(\nu)}=\oplus_{\nu'\subseteq\nu}\cal A^{\nu'}\subset \cal A$.
First of all let us define a {\em sector-reduction}, or $s$-reduction of
an element $f\in\cal A^{(\nu)}$ modulo a basis of the ideal
$\cal I^\nu=I\cap\cal A^\nu$.
Take the proper form of $f$ and let
$f^\nu$ be the sum of the terms in this decomposition that lie in
$\cal A^\nu$.
If $f^\nu$ is equal to zero the $s$-reduction stops.
Otherwise we look for a monomial $g\in\cal A^{(\nu)}$ and
a coefficient $c\in \cal K$ such that
the degrees of $f$ and $gf_i$ for some element of the bases $f_i$
coincide, that $(c f-gf_i)^\nu$ is zero or its degree is smaller
and that the value of $c$ at the point $(a_1,\ldots,a_n)$ is non-zero,
where $a_i=1$ if $i\in\nu$ and $0$ otherwise.
The procedure is repeated while possible.

A {\em sector} basis, or, an $s$-basis for a sector $\sg_{\nu}$ is a
basis of the ideal $I^\nu$ such that
the number of possible degrees
$f^\nu$, where $f$ is the result of the $s$-reduction,
is finite.
Such a basis provides the possibility
of a reduction to master integrals \textit{and} integrals whose
indices lie in \textit{lower} sectors, i.e. $\sg_{\nu'}$ for $\nu'\subset\nu$.

To prove that an $s$-basis always exists
is an open problem, but in all our
examples they do exist, and it turns out that in all known
examples where one can construct a Gr\"obner basis it is an
$s$-basis as well, although this does not follow from the
definition.

If $\nu=\emptyset$ then
an $s$-basis is a Gr\"obner basis
% definitions of a Gr\"obner basis and
%an $s$-basis coincide,
but generally it is not.
%an $s$-bases is not
%a Gr\"obner basis.
Since the sector $\sg_\emptyset$ is trivial, we do not have to
construct a single Gr\"obner bases. Still, it is most complicated
to construct $s$-bases for minimal sectors
(a sector $\sg_{\nu}$ is said to be \textit{minimal} if it is non-trivial but
all lower sectors are trivial).

Having constructed $s$-bases for all non-trivial sectors we have an
algorithm to evaluate $F$ at any point. Indeed, we
choose a sector containing the point we need, run the $s$-reduction
algorithm
for this sector, expressing $F$ in terms of some master integrals
and values for lowers sectors, then repeat the procedure for all
those sectors. Eventually we reduce $F$ to the master integrals.

The Buchberger algorithm leads us to
constructing a Gr\"obner basis that is hopefully an $s$-basis,
but this has no use for us since this does not simplify anything.
The second important point is that the Buchberger algorithm can be terminated
%at a certain point --- the point
when the Gr\"obner basis is not
yet constructed but the `current' basis  %at this point
already provides us the $s$-reduction,
so that it is an $s$-basis (we have criteria that show
whether a basis is an $s$-basis).

Let us illustrate how this idea works on the same example.
The initial basis turns out to be an $s$-basis.
First, let us observe that the degree of $f'_1$ is $(1,1)$
and the leading  coefficient is $q^2 a_2$, i.e. is a non-zero function
in the positive sector, hence we are capable of
making reduction steps if the highest degree of an element being
reduced is different from $(l,0)$ and $(0,l)$.
Now let $f$ be a polynomial whose highest degree is
$(l,0)$ and the leading coefficient is $c$. Then
\be
f'=q^2 (l-1+a_1)f + c Y^{l-2}_1 Y_2^{-1} f'_2
\ee
is an element of $\cal A$.
Let us take the proper form of $f$
(implying that the numbers $d_{i,j}$ can be now negative)
and calculate its highest degree without paying attention to the
terms with negative $d_{i,j}$. Obviously it is smaller than the
degree of $f$.
Now if we take the value of $f'$ at $(1,1)$ we will have the
elements like $F(j,0)$ among others. But the boundary conditions
state that they are equal to zero, so that we have nothing to worry
about. Now we can move further and reduce $f'$ as we did before.
Finally we get
\be
f=X+c(A_1,A_2)+Y
\ee
where $X\in {\cal I}$, $c$ is a rational function
and $Y$ is an element of $\cal A$ such that in its proper form
there are no degrees ${d_1,d_2}$ with $d_1\geq 0$ and $d_2\geq 0$.
Taking the value at $(1,1)$ we obtain the reduction of any value
of $F$ to the value of the master integral $F(1,1)$.
(See also \cite{GeRo}.)

%When constructing it, we shall stop the procedure once it provides
%a reduction to master integrals.

\section{Examples}

Let us consider a modification of Example~2;
now we have a non-zero mass $m_1=m$:

{\em Example~3.} Propagator integrals with the masses $m$ and $0$,
\be
F(a_1,a_2) =
\int \frac{\dd^d k}{(k^{2} -m^{2})^{a_1} [(q-k)^{2}]^{a_2}} \;.
\label{fi-l-gen}
\ee
The integrals are zero if $a_1\leq 0$.
The corresponding IBP relations generate the following elements:
\bea
f_1&=&
d - 2 a_1 - a_2 - 2 m^2 a_1 Y_1 - m^2 a_2 Y_2 + q^2 a_2 Y_2 -
  a_2 Y_2 Y_1^{-1}
\nn \\
f_2&=&
a_2 -a_1  - m^2 a_1 Y_1 - q^2 a_1 Y_1 - m^2 a_2 Y_2 + q^2 a_2 Y_2 -
  a_2 Y_2 Y_1^{-1} + a_1 Y_1 Y_2^{-1} \,. \nn
\eea
We have to consider two sectors, $\sg_{\{1,2\}}$ and $\sg_{\{1\}}$.

Using the lexicographical ordering,
we obtain, for the sector $\sg_{\{1,2\}}$, the $s$-basis consisting of
two elements:
\bea
g_{11}&=&
Y_1 ^2 + a_1  Y_1 ^2 + 3 Y_1  Y_2  - d Y_1  Y_2  +
  a_1  Y_1  Y_2  + 2 a_2  Y_1  Y_2  + m^2  Y_1 ^2 Y_2
\nn \\ &&
- q^2  Y_1 ^2 Y_2  + m^2  a_1  Y_1 ^2 Y_2  - q^2  a_1  Y_1 ^2 Y_2
\;, \nn
\\
g_{12}&=&
 -3 Y_1  Y_2  + d Y_1  Y_2  - 2 a_1  Y_1  Y_2  - a_2  Y_1  Y_2  -
  2 m^2  Y_1 ^2 Y_2  - 2 m^2  a_1  Y_1 ^2 Y_2  - Y_2 ^2
\nn \\ &&
-  a_2  Y_2 ^2 - m^2  Y_1  Y_2 ^2 + q^2  Y_1  Y_2 ^2
- m^2  a_2  Y_1  Y_2 ^2 + q^2  a_2  Y_1  Y_2 ^2
\,. \nn
\eea
For the sector $\sg_{\{1\}}$, we obtain the following $s$-basis:
\bea
g_{21}&=&
1 - a_2  + m^2  Y_1  - q^2  Y_1  - m^2  a_2  Y_1  + q^2  a_2  Y_1
- Y_1  Y_2^{-1}
+ d Y_1  Y_2^{-1}   - 2 a_1  Y_1  Y_2^{-1}
\nn \\ &&
-  a_2  Y_1  Y_2^{-1}   - 2 m^2  Y_1 ^2 Y_2^{-1}   - 2 m^2  a_1  Y_1 ^2 Y_2^{-1}
\;, \nn
\\
g_{22}&=&
 -2 m^2  + 2 m^2  a_2  - 2 m^4 Y_1  + 2 m^2  q^2  Y_1  +
  2 m^4 a_2  Y_1  - 2 m^2  q^2  a_2  Y_1  - 2 Y_2^{-1}   + a_2  Y_2^{-1}
\nn \\ &&
+ 2 m^2  Y_1  Y_2^{-1}
  + 2 q^2  Y_1  Y_2^{-1}   + 2 m^2  a_1  Y_1  Y_2^{-1}
-  m^2  a_2  Y_1  Y_2^{-1}   - q^2  a_2  Y_1  Y_2^{-1}
\nn \\ &&
+ 2 m^4 Y_1 ^2 Y_2^{-1}
+  2 m^2  q^2  Y_1 ^2 Y_2^{-1}
  + 2 m^4 a_1  Y_1 ^2 Y_2^{-1}   +
  2 m^2  q^2  a_1  Y_1 ^2 Y_2^{-1}
\nn \\ &&
- d Y_1  Y_2^{-2} +
  2 a_1  Y_1  Y_2^{-2} + a_2  Y_1  Y_2^{-2}
\,. \nn
\eea
The reduction based on the two constructed $s$-sectors
reveals two master integrals, $F(1,1)$ and $F(1,0)$, in accordance with
results obtained by other ways. (See, e.g., Chapters~5 and~6 of \cite{EFI}
and \cite{GeRo}.)

{\em Example~4.} Two-loop massless propagator diagram of Fig.~1.
\begin {figure} [htbp]
\begin{picture}(100,100)(-150,-50)
\Line(2,0)(20,0)
\Line(80,0)(98,0)
\Line(50,30)(50,-30)
\CArc(50,0)(30,0,180)
\CArc(50,0)(30,180,360)
\Text(24,-30)[]{\small $2$}
\Text(24,30)[]{\small $1$}
\Text(76,-30)[]{\small $4$}
\Text(76,30)[]{\small $3$}
\Text(58,0)[]{\small $5$}
\Vertex(20,0){1.5}
\Vertex(80,0){1.5}
\Vertex(50,30){1.5}
\Vertex(50,-30){1.5}
\end{picture}
\caption {Two-loop propagator diagram}
\end{figure}
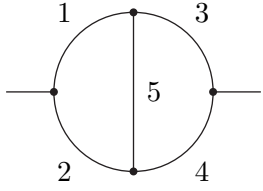

The corresponding family of Feynman integrals is
\be
F(a_1,a_2,a_3,a_4,a_5) =
\int\int \frac{\dd^dk\, \dd^dl}{(k^2)^{a_1} [(q-k)^2]^{a_2}
(l^2)^{a_3} [(q-l)^2]^{a_4}[(k-l)^2]^{a_5}} \; .
\label{two-loop-se}
\ee
There are boundary conditions which correspond to setting to zero
integrals without scale:
$F(a_1,a_2,a_3,a_4,a_5)=0$ , if $a_i,a_5\leq 0$ for $i=1,\ldots,4,$
or $a_1,a_2\leq 0$, or $a_3,a_4\leq 0$, or $a_1,a_3\leq 0$,
or $a_2,a_4\leq 0$.
The integrals are symmetrical:
\[
{}F(a_1,a_2,a_3,a_4,a_5)={}F(a_2,a_1,a_4,a_3,a_5)
= F(a_3,a_4,a_1,a_2,a_5)\,.
\]
The corresponding IBP relations generate the following elements:
\bea
f_1&=&(d-2a_1-a_2-a_5)
+a_2 Y_2(q^2- Y_1^{-1})
-a_5 Y_5  (Y_1^{-1} -Y_3^{-1} )\,,
\nn\\
f_2&=&(d-a_2-2a_3-a_5)
+a_4 Y_4 (q^2- Y_3^{-1})
-a_5 Y_5 (Y_3^{-1} -Y_1^{-1})\,,
\nn\\
f_3&=&(d-a_1-a_2-2a_5)
+a_1 Y_1 (Y_3^{-1} -Y_5^{-1})
+a_2 Y_2 (Y_4^{-1} -Y_5^{-1})\,,
\nn\\
f_4&=&(d-a_3-a_4-2a_5)
+a_3 Y_3 (Y_1^{-1} -Y_5^{-1} )
+a_4 Y_4 (Y_2^{-1} -Y_5^{-1})\,,
\nn\\
f_5&=&(d-a_1-2a_2-a_5)
+a_1 Y_1  (q^2-Y_2^{-1})
-a_5 Y_5 (Y_2^{-1} -Y_4^{-1})\,,
\nn\\
f_6&=&(d-a_3-2a_4-a_5)
+a_3 Y_3  (q^2-Y_4^{-1})
-a_5 Y_5  (Y_4^{-1}-Y_2^{-1} )\,.\nn
\eea

As is well known, any integral of this class can be reduced,
due to IBP relations, in a very simple way, to integrals where at
least one of the indices is non-positive. Such integrals can be
evaluated recursively in terms of gamma functions using the
one-loop integration formula (\ref{ex52-res}). Let us point out that physicists
often stop the reduction whenever they arrive at integrals
expressed in terms of gamma functions. Imagine, however, that
we want to know the whole solution of the reduction procedure,
i.e. a reduction to a minimal number of the master integrals.
Then this example turns out be not so trivial and
provides a good possibility to test our algorithms.

This can be done in various ways. In our approach,
we apply our algorithm to construct $s$-bases corresponding to the
sectors $\sg_{\{1,2,3,4,5\}}$, $\sg_{\{2,3,4,5\}}$ as well
as three more (symmetrical)
sectors, the minimal sector $\sg_{\{1,2,3,4\}}$, the minimal sector
$\sg_{\{2,3,5\}}$ as well as one more (symmetrical) sector.
In the first three cases, we used the degree-lexicographical
ordering, and in the last case, some special ordering.
%the ordering corresponding
%to the linear combinations:
%$a_1+a_3,\;a_3,\;a_2+a_4,\;a_2,\;a_5$ turned out to be optimal.

For example, let us present the $s$-basis associated with
$\sg_{\{1234\}}$:
\bea
g_1&=& 2 Y_1  Y_2  Y_3  Y_5^{-1}  - a_5  Y_1  Y_2  Y_3  Y_5^{-1}
- 2 Y_1  Y_2  Y_4  Y_5^{-1}  +
  a_5  Y_1  Y_2  Y_4  Y_5^{-1}  - 2 Y_1  Y_3  Y_4  Y_5^{-1}
\nn \\ &&  %\hspace*{-14mm}
+ a_5  Y_1  Y_3  Y_4  Y_5^{-1}  + 2 Y_2  Y_3  Y_4  Y_5^{-1}  -
  a_5  Y_2  Y_3  Y_4  Y_5^{-1}  + 2 q^2  Y_1  Y_2  Y_3  Y_4  Y_5^{-1}
\nn \\ &&  %\hspace*{-14mm}
- q^2  a_5  Y_1  Y_2  Y_3  Y_4  Y_5^{-1} + 2 q^2  Y_1  Y_2  Y_4 ^2 Y_5^{-1}
- q^2  a_5  Y_1  Y_2  Y_4 ^2 Y_5^{-1}  - 2 q^2  Y_1  Y_3  Y_4 ^2 Y_5^{-1}
\nn \\ &&  %\hspace*{-14mm}
+  q^2  a_5  Y_1  Y_3  Y_4 ^2 Y_5^{-1}  - 2 q^2  Y_2  Y_3  Y_4 ^2 Y_5^{-1}
+ q^2  a_5  Y_2  Y_3  Y_4 ^2 Y_5^{-1}  - (q^2)^2  Y_1  Y_2  Y_3  Y_4 ^2 Y_5^{-1}
  \nn \\ &&  %\hspace*{-14mm}
+  d (q^2)^2  Y_1  Y_2  Y_3  Y_4 ^2 Y_5^{-1}
  - (q^2)^2  a_3  Y_1  Y_2  Y_3  Y_4 ^2 Y_5^{-1}
- (q^2)^2  a_4  Y_1  Y_2  Y_3  Y_4 ^2 Y_5^{-1}
\nn \\ &&  %\hspace*{-14mm}
- 2 (q^2)^2  a_5  Y_1  Y_2  Y_3  Y_4 ^2  Y_5^{-1}
+ (q^2)^2  Y_2  Y_3 ^2 Y_4 ^2 Y_5^{-1}
+ (q^2)^2  a_3  Y_2  Y_3 ^2 Y_4 ^2  Y_5^{-1}
\nn \\ &&  %\hspace*{-14mm}
+ 2 (q^2)^2  Y_1  Y_3  Y_4 ^3 Y_5^{-1}
   + (q^2)^2  a_4  Y_1  Y_3  Y_4 ^3 Y_5^{-1}
- Y_1  Y_2  Y_3 ^2 Y_5^{-2}  - a_3  Y_1  Y_2  Y_3 ^2 Y_5^{-2}
\nn \\ &&  %\hspace*{-14mm}
+ a_3  Y_1  Y_2  Y_3  Y_4  Y_5^{-2}  - a_4  Y_1  Y_2  Y_3  Y_4  Y_5^{-2}
+  Y_1  Y_2  Y_4 ^2 Y_5^{-2} + a_4  Y_1  Y_2  Y_4 ^2 Y_5^{-2}
\nn \\ &&  %\hspace*{-14mm}
 - 6 q^2  Y_1  Y_2  Y_3  Y_4 ^2 Y_5^{-2}
 + 2 d q^2  Y_1  Y_2  Y_3  Y_4 ^2 Y_5^{-2}
- 3 q^2  a_3  Y_1  Y_2  Y_3  Y_4 ^2 Y_5^{-2}
\nn \\ &&  %\hspace*{-14mm}
- 4 q^2  a_4  Y_1  Y_2  Y_3  Y_4 ^2 Y_5^{-2}
- 2 q^2  a_5  Y_1  Y_2  Y_3  Y_4 ^2 Y_5^{-2}
- 2 q^2  Y_1  Y_2  Y_4 ^3 Y_5^{-2}
\nn \\ &&  %\hspace*{-14mm}
- q^2  a_4  Y_1  Y_2  Y_4 ^3 Y_5^{-2} \;,
\nn \\
g_2&=&
 Y_1  Y_2  Y_4 ^2 - a_5  Y_1  Y_2  Y_4 ^2 - Y_2  Y_3  Y_4 ^2 +
  a_5  Y_2  Y_3  Y_4 ^2 - 3 Y_1  Y_2  Y_3  Y_4 ^2 Y_5^{-1}
\nn \\ &&  %\hspace*{-14mm}
+ d Y_1  Y_2  Y_3  Y_4 ^2 Y_5^{-1}  - 2 a_3  Y_1  Y_2  Y_3  Y_4 ^2 Y_5^{-1}  -
  a_4  Y_1  Y_2  Y_3  Y_4 ^2 Y_5^{-1}  - a_5  Y_1  Y_2  Y_3  Y_4 ^2 Y_5^{-1}
\nn \\ &&  %\hspace*{-14mm}
- 2 Y_1  Y_2  Y_4 ^3 Y_5^{-1}
- a_4  Y_1  Y_2  Y_4 ^3 Y_5^{-1}  +
  2 q^2  Y_1  Y_2  Y_3  Y_4 ^3 Y_5^{-1}  + q^2  a_4  Y_1  Y_2  Y_3  Y_4 ^3 Y_5^{-1}
  \;,
\nn \\
g_3&=&
 -Y_1  Y_2  Y_3  Y_4  + a_5  Y_1  Y_2  Y_3  Y_4  + Y_1  Y_2  Y_4 ^2 -
  a_5  Y_1  Y_2  Y_4 ^2 + Y_1  Y_3  Y_4 ^2 - a_5  Y_1  Y_3  Y_4 ^2
\nn \\ &&  %\hspace*{-14mm}
- Y_2  Y_3  Y_4 ^2 + a_5  Y_2  Y_3  Y_4 ^2 + Y_1  Y_2  Y_3 ^2 Y_4  Y_5^{-1}
+ a_3  Y_1  Y_2  Y_3 ^2 Y_4  Y_5^{-1}  + Y_1  Y_2  Y_3  Y_4 ^2 Y_5^{-1}
\nn \\ &&  %\hspace*{-14mm}
-  a_3  Y_1  Y_2  Y_3  Y_4 ^2 Y_5^{-1}  + a_4  Y_1  Y_2  Y_3  Y_4 ^2 Y_5^{-1}
- q^2  Y_1  Y_2  Y_3 ^2 Y_4 ^2 Y_5^{-1}  - q^2  a_3  Y_1  Y_2  Y_3 ^2 Y_4 ^2 Y_5^{-1}
\nn \\ &&  %\hspace*{-14mm}
-  2 Y_1  Y_2  Y_4 ^3 Y_5^{-1}
- a_4  Y_1  Y_2  Y_4 ^3 Y_5^{-1}
+ 2 q^2  Y_1  Y_2  Y_3  Y_4 ^3 Y_5^{-1}  + q^2  a_4  Y_1  Y_2  Y_3  Y_4 ^3 Y_5^{-1}
\;,
\nn \\
g_4&=&
 -Y_1  Y_2  Y_4^2 + a_5  Y_1  Y_2  Y_4 ^2 + Y_2  Y_3  Y_4 ^2 -
  a_5  Y_2  Y_3  Y_4 ^2 - 2 Y_1  Y_2  Y_3  Y_4 ^2 Y_5^{-1}
\nn \\ &&  %\hspace*{-14mm}
+ d Y_1  Y_2  Y_3  Y_4 ^2 Y_5^{-1}  - 2 a_1  Y_1  Y_2  Y_3  Y_4 ^2 Y_5^{-1}  -
  a_2  Y_1  Y_2  Y_3  Y_4 ^2 Y_5^{-1}  - a_5  Y_1  Y_2  Y_3  Y_4 ^2 Y_5^{-1}
\nn \\ &&  %\hspace*{-14mm}
- Y_2 ^2 Y_3  Y_4 ^2 Y_5^{-1} - a_2  Y_2 ^2 Y_3  Y_4 ^2 Y_5^{-1}
+ q^2  Y_1  Y_2 ^2 Y_3  Y_4 ^2 Y_5^{-1}  + q^2  a_2  Y_1  Y_2 ^2 Y_3  Y_4 ^2 Y_5^{-1}
\;,
\nn \\
g_5&=&
Y_1  Y_2  Y_3  Y_4  - a_5  Y_1  Y_2  Y_3  Y_4  - Y_1  Y_2  Y_4 ^2 +
  a_5  Y_1  Y_2  Y_4 ^2 - Y_1  Y_3  Y_4 ^2
+ a_5  Y_1  Y_3  Y_4 ^2
\nn \\ &&  %\hspace*{-14mm}
+ Y_2  Y_3  Y_4 ^2 - a_5  Y_2  Y_3  Y_4 ^2 + Y_1 ^2 Y_3  Y_4 ^2 Y_5^{-1}  +
  a_1  Y_1 ^2 Y_3  Y_4 ^2 Y_5^{-1}
- a_1  Y_1  Y_2  Y_3  Y_4 ^2 Y_5^{-1}
\nn \\ &&  %\hspace*{-14mm}
+  a_2  Y_1  Y_2  Y_3  Y_4 ^2 Y_5^{-1}
  - q^2  Y_1 ^2 Y_2  Y_3  Y_4 ^2 Y_5^{-1}  -
  q^2  a_1  Y_1 ^2 Y_2  Y_3  Y_4 ^2 Y_5^{-1}  - Y_2 ^2 Y_3  Y_4 ^2 Y_5^{-1}
\nn \\ &&  %\hspace*{-14mm}
- a_2  Y_2 ^2 Y_3  Y_4 ^2 Y_5^{-1}  + q^2  Y_1  Y_2 ^2 Y_3  Y_4 ^2 Y_5^{-1}  +
  q^2  a_2  Y_1  Y_2 ^2 Y_3  Y_4 ^2 Y_5^{-1}
\;,
\nn\\
g_6&=&
 -q^2  Y_1  Y_2  Y_4 ^2 +
  q^2  a_5  Y_1  Y_2  Y_4 ^2 + q^2  Y_2  Y_3  Y_4 ^2 - q^2  a_5  Y_2  Y_3  Y_4 ^2 +
  2 Y_1  Y_2  Y_3  Y_4  Y_5^{-1}
\nn \\ &&  %\hspace*{-14mm}
- a_5  Y_1  Y_2  Y_3  Y_4  Y_5^{-1}  -
  2 Y_1  Y_2  Y_4 ^2 Y_5^{-1}  + a_5  Y_1  Y_2  Y_4 ^2 Y_5^{-1}  -
  2 Y_1  Y_3  Y_4 ^2 Y_5^{-1}  + a_5  Y_1  Y_3  Y_4 ^2 Y_5^{-1}
\nn \\ &&  %\hspace*{-14mm}
+ 2 Y_2  Y_3  Y_4 ^2 Y_5^{-1}  - a_5  Y_2  Y_3  Y_4 ^2 Y_5^{-1}  +
  2 q^2  Y_1  Y_2  Y_3  Y_4 ^2 Y_5^{-1}  + q^2  a_3  Y_1  Y_2  Y_3  Y_4 ^2 Y_5^{-1}
\nn \\ &&  %\hspace*{-14mm}
-  q^2  a_5  Y_1  Y_2  Y_3  Y_4 ^2 Y_5^{-1}  + q^2  Y_2  Y_3 ^2 Y_4 ^2 Y_5^{-1}  +
  q^2  a_3  Y_2  Y_3 ^2 Y_4 ^2 Y_5^{-1}  + 2 q^2  Y_1  Y_2  Y_4 ^3 Y_5^{-1}
\nn \\ &&  %\hspace*{-14mm}
+  q^2  a_4  Y_1  Y_2  Y_4 ^3 Y_5^{-1}  + 2 q^2  Y_1  Y_3  Y_4 ^3 Y_5^{-1}  +
  q^2  a_4  Y_1  Y_3  Y_4 ^3 Y_5^{-1}  - 2 (q^2)^2  Y_1  Y_2  Y_3  Y_4 ^3 Y_5^{-1}
\nn \\ &&  %\hspace*{-14mm}
- (q^2)^2  a_4  Y_1  Y_2  Y_3  Y_4 ^3 Y_5^{-1}  - Y_1  Y_2  Y_3 ^2 Y_4  Y_5^{-2}  -
  a_3  Y_1  Y_2  Y_3 ^2 Y_4  Y_5^{-2}  - Y_1  Y_2  Y_3  Y_4 ^2 Y_5^{-2}
\nn \\ &&  %\hspace*{-14mm}
+ a_3  Y_1  Y_2  Y_3  Y_4 ^2 Y_5^{-2}  - a_4  Y_1  Y_2  Y_3  Y_4 ^2 Y_5^{-2}  +
  2 Y_1  Y_2  Y_4 ^3 Y_5^{-2}  + a_4  Y_1  Y_2  Y_4 ^3 Y_5^{-2}
\nn \\ &&  %\hspace*{-14mm}
- 4 q^2  Y_1  Y_2  Y_3  Y_4 ^3 Y_5^{-2}  - 2 q^2  a_4  Y_1  Y_2  Y_3  Y_4 ^3 Y_5^{-2}
\;,
\nn\\
g_7&=&
  q^2  Y_1  Y_2  Y_4 ^2 - q^2  a_5  Y_1  Y_2  Y_4 ^2 -
  q^2  Y_2  Y_3  Y_4 ^2 + q^2  a_5  Y_2  Y_3  Y_4 ^2 -
  2 Y_1  Y_2  Y_3  Y_4  Y_5^{-1}
\nn \\ &&
+ a_5  Y_1  Y_2  Y_3  Y_4  Y_5^{-1}  +
  q^2  Y_1  Y_2 ^2 Y_3  Y_4  Y_5^{-1}  + q^2  a_2  Y_1  Y_2 ^2 Y_3  Y_4  Y_5^{-1}  +
  2 Y_1  Y_2  Y_4 ^2 Y_5^{-1}
\nn \\ &&
- a_5  Y_1  Y_2  Y_4 ^2 Y_5^{-1}  +
  q^2  Y_1 ^2 Y_2  Y_4 ^2 Y_5^{-1}  + q^2  a_1  Y_1 ^2 Y_2  Y_4 ^2 Y_5^{-1}  +
  2 Y_1  Y_3  Y_4 ^2 Y_5^{-1}
\nn \\ &&
- a_5  Y_1  Y_3  Y_4 ^2 Y_5^{-1}  -
  2 Y_2  Y_3  Y_4 ^2 Y_5^{-1}  + a_5  Y_2  Y_3  Y_4 ^2 Y_5^{-1}  +
  2 q^2  Y_1  Y_2  Y_3  Y_4 ^2 Y_5^{-1}
\nn \\ &&
+ q^2  a_1  Y_1  Y_2  Y_3  Y_4 ^2 Y_5^{-1}  -
  q^2  a_5  Y_1  Y_2  Y_3  Y_4 ^2 Y_5^{-1}  + q^2  Y_2 ^2 Y_3  Y_4 ^2 Y_5^{-1}  +
  q^2  a_2  Y_2 ^2 Y_3  Y_4 ^2 Y_5^{-1}
\nn \\ &&
- (q^2)^2  Y_1  Y_2 ^2 Y_3  Y_4 ^2 Y_5^{-1}  -
  (q^2)^2  a_2  Y_1  Y_2 ^2 Y_3  Y_4 ^2 Y_5^{-1}  - Y_1 ^2 Y_3  Y_4 ^2 Y_5^{-2}  -
  a_1  Y_1 ^2 Y_3  Y_4 ^2 Y_5^{-2}
\nn \\ &&
+ a_1  Y_1  Y_2  Y_3  Y_4 ^2 Y_5^{-2}  -
  a_2  Y_1  Y_2  Y_3  Y_4 ^2 Y_5^{-2}  + Y_2 ^2 Y_3  Y_4 ^2 Y_5^{-2}
\nn \\ &&
+  a_2  Y_2 ^2 Y_3  Y_4 ^2 Y_5^{-2}  - 2 q^2  Y_1  Y_2 ^2 Y_3  Y_4 ^2 Y_5^{-2}  -
  2 q^2  a_2  Y_1  Y_2 ^2 Y_3  Y_4 ^2 Y_5^{-2}
\;. \nn
\eea
The reduction based on the constructed $s$-sectors
reveals three master integrals, $F(1,1,1,1,0)$, $F(0,1,1,0,1)$
and $F(1,0,0,1,1)$ (the last two of them are equal because
of the symmetry), in accordance with results obtained
by other ways. (See, e.g., Chapters~5 and~6 of \cite{EFI}.)

%%%%%%%%%%%%%%%%%%%%%%%%%%%%
Our last example is

{\em Example~5.} Two-loop Feynman integrals for the heavy quark static potential
corresponding to Fig.~2 with $v\cdot q=0.$
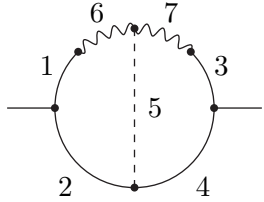
\begin {figure} [htbp]
\begin{picture}(100,100)(-150,-50)
\Line(2,0)(20,0)
\Line(80,0)(98,0)
\DashLine(50,30)(50,-30){3}
\CArc(50,0)(30,0,45)
\PhotonArc(50,0)(30,45,90){2}{4}
\PhotonArc(50,0)(30,90,135){2}{4}
\CArc(50,0)(30,135,360)

\Text(17,16)[]{\small $1$}
\Text(83,16)[]{\small $3$}
\Text(24,-30)[]{\small $2$}
\Text(76,-30)[]{\small $4$}
\Text(58,0)[]{\small $5$}
\Text(36,36)[]{\small $6$}
\Text(64,36)[]{\small $7$}

\Vertex(20,0){1.5}
\Vertex(80,0){1.5}
\Vertex(50,30){1.5}
\Vertex(50,-30){1.5}
\Vertex(28.7867965644,21.2132034){1.5}
\Vertex(71.213203436,21.2132034){1.5}
\end{picture}
\caption {Feynman diagram contributing to the three-loop static
quark potential. A wavy line denotes a propagator for the static source
and the dotted line denotes the scalar propagator with the index shifted
by $\ep$.}
\end{figure}

We shall consider diagrams contributing to the three-loop
static potential corresponding to Fig.~2.
They are obtained from the corresponding two-loop diagrams by
inserting a one-loop diagram into the central line.
Indeed, the integration over the loop-momentum of
the insertion can be performed explicitly, by means of
(\ref{ex52-res}), and one obtains,
up to a factor expressed in terms of gamma functions, Feynman
integrals of Fig.~2, where the index of the central
line\footnote{
A more general case, with $a_5\to a_5+ r \ep$ and integer $r$,
relevant to $r$-loop massless insertions can be considered
on the same footing.}
is $a_5+\ep\equiv a_5+(4-d)/2$ with integer $a_5$.
So, we arrive at the following family of integrals:\footnote{
These integrals with the integer index of the central line
contribute to the two-loop static quark potential and were
calculated in \cite{PSKPSS,ST}.}
\bea
 F(a_1,a_2,a_3,a_4,a_5,a_6,a_7) &=&
  \int\int\frac{{\rm d}^dk{\rm d}^dl}{(-k^2)^{a_1}
 [-(k-q)^2]^{a_2}(-l^2)^{a_3}[-(l-q)^2]^{a_4} }
    \nn \\ &&
  \times \frac{1}{ [-(k-l)^2]^{a_5+\ep}(-v \cdot k)^{a_6}(-v \cdot l)^{a_7}}\,.
\eea
We have turned to the $-k^2$ dependence of the propagators because
this choice is more natural when at least one index, $a_5+\ep$ is
not integer.
The integrals are symmetrical:
\[
{}F(a_1,a_2,a_3,a_4,a_5,a_6,a_7)={}F(a_2,a_1,a_4,a_3,a_5,a_6,a_7)
= F(a_3,a_4,a_1,a_2,a_5,a_7,a_6)\,.
\]
They are equal to zero, if $a_1,a_3\leq 0$, or $a_2,a_4\leq 0$, or
$a_1,a_2,a_6\leq 0$, or $a_3,a_4,a_7\leq 0$.

The IBP relations generate the following elements
in the case where $a_5$ is not shifted by $\ep$ (i.e. the case of
the diagrams relevant to the two-loop static quark potential
considered in \cite{PSKPSS,ST})
\begin{eqnarray}
f_1&=&(d-2a_1-a_2-a_5-a_6)
-a_2 Y_2  (q^2+ Y_1^{-1})
-a_5 Y_5 (Y_1^{-1} -Y_3^{-1})\,,
\nn \\
f_2&=&(d-a_2-2a_3-a_5-a_7)
-a_4 Y_4  (q^2+Y_3^{-1} )
-a_5 Y_5  (Y_3^{-1}-Y_1^{-1} )\,,
\nn \\
f_3&=&(d-a_1-a_2-2a_5-a_6 )
+a_1 Y_1 (Y_3^{-1} -Y_5^{-1} )
+a_2 Y_2  (Y_4^{-1}-Y_5^{-1})
+a_6 Y_6 Y_7^{-1} \,,
\nn \\
f_4&=&(d-a_3-a_4-2a_5-a_7 )
+a_3 Y_3  (Y_1^{-1} -Y_5^{-1})
+a_4 Y_4  (Y_2^{-1} -Y_5^{-1})
+a_7 Y_6^{-1} Y_7  \,,
\nn \\
f_5&=&(d-a_1-2a_2-a_5-a_6)
-a_1 Y_1 (q^2+Y_2^{-1})
-a_5 Y_5 (Y_2^{-1}-Y_4^{-1})\,,
\nn \\
f_6&=&(d-a_3-2a_4-a_5-a_7)
-a_3 Y_3 (q^2+ Y_4^{-1})
-a_5 Y_5 (Y_4^{-1}-Y_2^{-1})\,,
\nn \\
f_7&=& 2a_1 Y_1 Y_6^{-1}
+2a_2 Y_2 Y_6^{-1}
+a_5 Y_5 (Y_6^{-1}- Y_7^{-1})
-v^2 a_6 Y_6 \,,
\nn \\
f_8&=& 2a_3 Y_3 Y_7^{-1}
+2a_4 Y_4 Y_7^{-1}
-a_5 Y_5 (Y_6^{-1}- Y_7^{-1})
-v^2 a_7 Y_7  \,. \nn
\end{eqnarray}
So, the IBP elements we need are obtained from these by replacing
$a_5$ with $a_5+\ep$.

Our algorithm works successfully in this example and gives us a
family of $s$-bases which provide the possibility
of a reduction to master integrals. The elements of the bases are
rather lengthy, typically, with hundreds of terms, so that we do
not present them in this short paper.
These $s$-bases correspond to the following sectors:
$\sg_{\{1,2,3,4,5,6,7\}}$, $\sg_{\{2,3,4,5,6,7\}}$, $\sg_{\{1,2,3,4,5,7\}}$,
$\sg_{\{3,4,5,6,7\}}$, $\sg_{\{2,3,5,6,7\}}$, $\sg_{\{2,3,4,5,7\}}$,
$\sg_{\{2,3,4,5,6\}}$, $\sg_{\{1,2,3,4,5\}}$,  $\sg_{\{2,3,4,5\}}$,
$\sg_{\{2,3,5,6\}}$ and other sectors obtained by the symmetry transformations.

We obtain the %following
master integrals:
$I_1 = F(1,1,1,1,0,1,1), I_{21} = F(1,1,1,1,0,0,1),$ $I_{22} =
F(1,1,1,1,0,1,0),
I_3 = F(1,1,1,1,0,0,0)\,.$
We have $I_{21}=I_{12}=I_2$ because of the symmetry.
We also obtain
$I_{51} = F (1,0,0,1,1,1,1), I_{71} = F (1,0,0,1,1,0,1),
    I_{81} = F (1,0,0,1,1,1,0),
    I_{41} = F (1,0,0,1,1,0,0).$
We have $I_{71}=I_{81}=I_7$ because of the symmetry.
Moreover, we have other copies,
$I_{52},I_{72},I_{82},I_{42}$,
 of this last family of the master
integrals which are obtained by the symmetry transformation
$(1\lra 2, 3\lra 4$).
We also obtain
$I_{61} = F (0,0,1,1,1,1,0),
 \bar{I}_{61} = F (0,0,1,1,1,2,0)$
as well as the corresponding symmetrical family.

To calculate the master integrals one can use the
threefold Mellin--Barnes representation
\bea
F(a_1,a_2,a_3,a_4,a_5,a_6,a_7) &=&
\frac{\left(\I\pi^{d/2} \right)^2 2^{a_7 - 1}(v^2)^{-a_{67}/2}}{
\prod_{l=3,4,5,7}\Gm(a_l) \Gm(4-a_{3457}-2\ep)
(Q^2)^{a_{12345}-4+2\ep+a_{67}/2} }
\nn \\ &&  \hspace*{-52mm}\times
\frac{1}{(2\pi \I)^3}
\int_{-\I\infty}^{+\I\infty}\ldots \int_{-\I\infty}^{+\I\infty}
\dd z_1 \dd z_2 \dd z_3
\frac{\Gm(a_{12345} + a_{67}/2   + 2\ep - 4 + z_3)}
{\Gm(a_{345} + a_{67}/2+ \ep - 3/2 + z_1 + z_2 + z_3)}
\nn \\ &&  \hspace*{-52mm}\times
\frac{\Gm(a_3 + z_1 + z_3)\Gm(a_4 + z_2 + z_3)
\Gm(a_{345} + a_7/2 + \ep - 2 + z_1 + z_2 + z_3)
}
{\Gm(a_1 - z_1)\Gm(a_2 - z_2)
\Gm(8 - a_{1267} - 2 a_{345} - 4\ep - z_1 - z_2 - 2z_3)}
\nn \\ &&  \hspace*{-52mm}\times
\Gm(a_{345} + a_7/2 + \ep - 3/2 + z_1 + z_2 + z_3)
\Gm(4 - 2 a_{34} - a_{57} - 2\ep - z_1 - z_2 - 2z_3)
\nn \\ &&  \hspace*{-52mm}\times
\Gm(4 - a_{1345} - a_{67}/2 - 2\ep - z_2 - z_3)
\Gm(2 - a_{345} - \ep - z_1 - z_2 -z_3)
\nn \\ &&  \hspace*{-52mm}\times
\Gm(4 - a_{2345} - a_{67}/2  - 2\ep - z_1 - z_3)
\Gm(-z_1)\Gm(-z_2)\Gm(-z_3)\;,
\label{3MB}
\eea
where  $a_{12345}=a_1+a_2+a_3+a_4+a_5$ etc.
The integrations over the variables $z_i$ go from $-\infty$ to
$+\infty$ in the complex plane. The contours are chosen in the
standard way: the poles of gamma functions with the $-z_i$
dependence are to the right of the contour and
the poles with the $+z_i$ dependence are to the left of it.
This representation can be derived by applying Feynman
parameterization to the subloop integral over $l$, introducing
then three MB integrations and, finally, integrating over $k$.
(See Chapter 4 of \cite{EFI} for details of this method,
%based on the MB representation,
with multiple examples.)

We obtain the following results for the master integrals:
\bea
I_1 &=& \frac{\left(i \pi^{d/2}e^{-\gm_{\rm E}\ep}\right)^2}{Q^{4+6\ep}v^2}
\left[-\frac{8 \pi^2}{9 \ep} -\frac{16 \pi^2}{9} + \frac{40 \zeta(3)}{3}
+O(\ep)\right]
 \;, \nn \\ %  \hspace*{-50mm}
I_2 &=& \frac{\left(i \pi^{d/2}e^{-\gm_{\rm E}\ep}\right)^2}{Q^{3+6\ep}v}
\left[ \frac{\pi^4}{3} +O(\ep)\right]
 \;, \nn \\ %  \hspace*{-50mm}
I_3 &=& \frac{\left(i \pi^{d/2}e^{-\gm_{\rm E}\ep}\right)^2}{Q^{2+6\ep}}
\left[6\zeta(3) + \left(\frac{\pi^4}{10} + 12\zeta(3)\right)\ep
+O(\ep^2)\right]
 \;, \nn \\ %  \hspace*{-50mm}
I_4 &=&  \left(i \pi^{d/2}\right)^2 Q^{2-6\ep}
\frac{\Gm\left(1 - 2 \ep\right) \Gm\left(1 - \ep\right)^2 \Gm\left(3 \ep - 1\right)}
{\Gm\left(3 - 4 \ep\right) \Gm\left(1 + \ep\right) }
\;, \nn \\ %&&  \hspace*{-50mm}
I_5 &=& \frac{\left(i \pi^{d/2}e^{-\gm_{\rm E}\ep}\right)^2}{Q^{6\ep-2}v^2}
\left[\frac{4\pi^2}{9 \ep} + \frac{32 \pi^2}{9}
- \frac{8\zeta(3)}{3}
+O(\ep)\right]
 \;, \nn \\ %  \hspace*{-50mm}
I_6 &=&  \left(i \pi^{d/2}\right)^2 Q^{2-6\ep}
\frac{4^{1 - 2\ep} \sqrt{\pi} \Gm\left(3/2 - 3 \ep\right)^2
\Gm\left(1 - 2 \ep\right)
\Gm\left(3 \ep - 1/2\right)\Gm\left(4\ep - 1\right) }
{\Gm\left(3 - 6 \ep\right) \Gm\left(2 \ep\right)\Gm\left(1 + \ep\right)  v}
\;, \nn \\ %&&  \hspace*{-50mm}
\bar{I}_6 &=&  \left(i \pi^{d/2}\right)^2 Q^{2-6\ep}
\frac{4^{1 - 2\ep} \sqrt{\pi} \Gm\left(1-3 \ep\right)^2 \Gm\left(1-2 \ep\right)
\Gm\left(3 \ep\right)\Gm\left(4 \ep\right) }
{\Gm\left(2-6 \ep\right) \Gm\left(1+\ep\right) \Gm\left(1/2+2 \ep\right)  v^2}
\;, \nn \\ %&&  \hspace*{-50mm}
I_7 &=&  \left(i \pi^{d/2}\right)^2
\frac{\sqrt{\pi} \Gm\left(3/2 - 3 \ep\right) \Gm\left(1 - 2\ep\right)
\Gm\left(1/2 - \ep\right) \Gm\left(1 - \ep\right)
\Gm\left(3 \ep - 1/2\right) }
{Q^{6\ep-1} v \Gm\left(2 - 4 \ep\right)  \Gm\left(2 - 3 \ep\right)
\Gm\left(1 + \ep\right)}
\;, \nn \\ %&&  \hspace*{-50mm}
I_8 &=& I_7
\;, \nn
\eea
where $Q=\sqrt{-q^2}$ and $v=\sqrt{v^2}$.

Observe that some of the integrals are expressed explicitly in
terms of gamma functions for general values of $\ep$ while results
for some other integrals are presented in expansion in a Laurent
series in $\ep$. The depth of this expansion can be made greater
whenever necessary.

For example, we obtain the following reductions to master
integrals by our algorithm:
\bea
F(1,1,1,1,1,1,-1)&=&
-\frac{2 Q^2 v^2}{(3 d - 10)} \bar{I}_2 - 3 I_3 -
 \frac{ 8 (d - 3) (2 d - 7) (11 d - 46)}{(d - 4)^2 (3 d - 14) Q^4} I_4
\nn \\ &&
+ \frac{4 (3 d - 11) (7 d - 30) v^2}
{(d - 4) (3 d - 14) (3 d - 10)Q^2} \bar{I}_6 \,,
\\
F(2,1,1,1,1,1,1)&=&
-\frac{3 d - 14}{2 Q^2} I_1
\nn \\ && \hspace*{-37mm}
- \frac{4 (d - 3) (d - 2) (2d - 7) (3 d -10) (9 d - 40)}
{(d - 5) (d - 4) (2d - 11) (3 d - 16) (3 d - 14) Q^8 v^2} I_4
\nn \\ && \hspace*{-37mm}
- \frac{3 (d - 4) (4 d - 17) (4 d - 15)}{(2 (d - 5) (2d - 11) Q^6} I_5
-  \frac{16 (3 d - 13) (3 d - 11)}
{(2 d - 11) (3 d - 16) (3 d - 14) Q^6} \bar{I}_6 \,,
\eea
which can be checked straightforwardly, by evaluating these
integrals, in expansion in $\ep$, using the MB representation
(\ref{3MB}).

\section{Conclusion}

We have developed an algorithm which is a generalization
of the Buchberger algorithm to the reduction problem for
Feynman integrals and modified it in such a way that it
works at the level of modern calculations.
We have described the main features of the algorithm.
For the examples considered, it works rather fast ---
these are seconds of CPU time for Example~4 and minutes
for Example~5, both for constructing $s$-bases and reduction to
master integrals.
In fact, it has turned out that our algorithm works successfully
even at a higher level, in a reduction problem with
nine indices \cite{GSS}.
Still to perform more sophisticated calculations, further
modifications and optimizations are needed.
One of possible ways to improve the algorithm is to
combine its basic points with that of algorithms based on
Janet bases \cite{Gerdt05}.
We hope to report on our progress in future publications.
We also postpone to solve various mathematical problems connected
with our algorithm.

\vspace{0.2 cm}

{\em Acknowledgments.}
We are grateful to I.V.~Arzhantsev, V.P.~Gerdt, Y.~Schr\"oder
and M.~Steinhauser  for helpful discussions and careful reading of
draft versions of the paper.
The work of A.S. was supported by the Russian Foundation for Basic
Research through  grant 05-01-00988 and
by CRDF through grant RM1-2543.
The work of V.S. was supported by the Russian Foundation for Basic
Research through grant 05-02-17645.

\end{document}